\documentclass[preprint,aps]{revtex4}

\usepackage{graphicx}
\usepackage{dcolumn}
\usepackage{bm}
\usepackage{times}

\begin{document}

\title{Sub-monolayer nucleation and growth of complex oxide heterostructures at high supersaturation and rapid flux modulation}
\author{M. Kareev$^{1}$, S. Prosandeev$^{1}$, Jian Liu$^{1}$, B. Gray$^{1}$, P. Ryan$^{2}$, A. Kareev$^{1}$, Eun Ju Moon$^{1}$, and J. Chakhalian$^{1}$}

\affiliation{$^{1}$Department of Physics, University of Arkansas, Fayetteville, AR 72701}
\affiliation{$^{2}$Advanced Photon Source, Argonne National Laboratory, Argonne, IL 60439}

\begin{abstract}
We report on the non-trivial nanoscale kinetics of the deposition of novel complex oxide  heterostructures  composed of a unit-cell thick correlated metal LaNiO$_{3}$  and dielectric LaAlO$_{3}$. The multilayers demonstrate exceptionally good crystallinity and surface morphology maintained over the large number of layers, as confirmed by AFM, RHEED, and synchrotron X-ray diffraction. To elucidate the physics behind the growth, the temperature of the substrate  and the deposition rate were varied over a wide range and the results were treated in the framework of a two-layer model. These results are of fundamental importance for synthesis of  new phases of complex oxide heterostructures.
\end{abstract}

\maketitle

Recently, novel many-body phenomena emerging at the oxide-oxide interface have attracted considerable attention motivated by the intriguing possibility that such interfacial phenomena can form the basis for new materials with greatly enhanced functionality\cite{ref1,ref2,ref3,ref4}. While the multiple and coupled degrees of freedom (lattice, spin, charge and orbital) potentially offer a means for manipulating and controlling novel functionalities, an understanding of the mechanisms of layer-by-layer (LBL)\ growth of such heterostructures (HSs)   with exquisite tunability and quality control  still present a formidable challenge. Naturally  the challenge of finding   parameters   for  LBL\ growth increases manyfold, when  the multilayers are subjected to large strain\cite{ref5}, broken translation symmetries of  the interface, and  dimensionality confined to a  single unit cell\cite{ref6}. During the past several years, extensive work on pulsed laser $\textit{interval}$ deposition combined with high pressure reflection high energy electron diffraction (HP RHEED) opened up new prospects for the stabilization of unusual material  phases\cite{ref7}.

Growing unit-cell thin complex oxide HSs in planar geometry is particularly challenging task. In this letter, we report on LBL growth of novel HSs having unit-cell-thick (uc) repetitions of correlated metal LaNiO$_{3}$ (LNO) interlaid with layers of the dielectric LaAlO$_{3}$ (LAO). The obtained results demonstrate that the growth kinetics of such 1uc LNO/1uc LAO superlattice (SL) has several interesting features leading to important fundamental conclusions applicable to growth of a wider range of complex-oxide HSs. Specifically, our findings elucidate that, in addition to temperature, modulated flux can be employed as a powerful kinetic handle for the fine-tuned manipulation of sub-monolayer growth regimes.

The LNO/LAO SLs have been grown on atomically flat TiO$_{2}$-terminated SrTiO$_{3}$(001) (STO) single-crystal substrates prepared by the recently developed wet-etch procedure\cite{ref8} to minimize surface and near-surface electronic defects. The PLD system was equipped with a newly developed high pressure RHEED system operating in the background O$_{2}$ pressure of up to 400 mTorr, with an advanced feedback control of the current and a custom developed 12-bit ultra-fast imaging system with a timing resolution $\leq$1 msec. We have employed real-time 3D imaging acquired per laser pulse to monitor both specular and diffuse intensities. By recording the temporal and thermal evolution of the diffracted intensity during the growth combined with \textit{ex-situ} atomic force microscopy (AFM), we were able to obtain exquisite details of the evolution of layer morphology.

At first, we demonstrate the difference between the \textit{continuous} low-frequency deposition (Figs. 1(a, b)) and interrupted high-frequency (Figs. (1c, d)) growth\cite{ref7,ref9} of the strongly correlated SLs. Figure 1(a) shows the result of a typical continuous deposition of $\sim$4 uc thick layer of LNO with an imposed slow growth rate defined by the laser frequency of 3 Hz. As seen, the RHEED specular intensity (RSI) exhibits strong damping right after a few monolayers attesting to the rapidly increasing surface disorder. The diffuse intensity (inset in Fig. 1(a)) shows a series of pronounced peaks typical of electrons transmitted through microcrystalline 3D-like islands\cite{ref10}. The large area 1$\mu$m $\times$ 1$\mu$m AFM scans shown in Fig. 1(b) further corroborate the presence of pronounced 3D microstructures. Figures 1(c, d) present the results obtained with the deposition rate increased up to 30 laser-pulses/sec and introduction of a prolonged time-delay (dwell time) up to 100 seconds between two consecutive unit cells.

As immediately evident from Figs. 1(c, d), such a LBL growth yields perfect coverage and an excellent morphology (see also auxiliary supplement\cite{ref11}). We further performed detailed temporal and temperature (T) dependent studies (from 450$^{o}$C to 780$^{o}$C) of the layer coverage during the growth of a sub-monolayer. The process of nucleation and condensation at high laser frequency revealed a non-trivial T-dependence. Specifically, the initial stage of deposition  (area 1 in Fig. 2(a)) is T-\textit{independent}, within the studied temperature interval, followed by two distinct non-linear T-dependences in the recovery stage 2 and the saturation stage 3. The origin of this interesting behavior can be connected to the specifics of the nucleation and condensation processes\cite{ref12,ref13} at  the limit of high supersaturation and flux modulation. Under the conditions in use, the high nucleation rate results in a {small critical nucleus radius\cite{ref14,ref15}, $i_{c}$ $\sim$1, which is on the order of a unit-cell. Secondly, we emphasize that a significant increase in the laser frequency pumps the ablated material into the vicinity of the substrate and thus further enlarges the supersaturation that scales with the mass density (for more data see auxiliary supplement\cite{ref11} and Ref.\cite{ref16}).

To corroborate the results deduced from the diffraction data, we have performed \textit{ex-situ} AFM imaging of the surface morphology by interrupting the growth at $\sim$0.3, 0.5 and 0.7 of layer coverage, $\theta$. The obtained morphological progression is shown in Figure 2(b). As seen in Figure 2(b-left), low layer coverage, $\sim$0.3$\theta$, (stage 1) is characterized by a large number of small-scale non-percolated islands which are characteristic of \textit{immobile} clusters with a large density. At the half coverage step, 0.5$\theta$, (Fig. 2(b-center)) a characteristic wavy pattern of the deposited material has been observed, which is the result of islands coalescing to form a low-dimensional percolated structure. This structure can possibly be described in the framework of the diffusion-limited aggregation theory\cite{ref14}. As the deposition continues (see Figure 2(b-right)), the growth approaches the layer completion, 0.7$\theta$, and looks morphologically imperfect implying that  the amount of delivered material is not sufficient to fill 'voids' in the under-layed layer.

To gain further insight into the kinetics of growth, we have calculated the surface coverage, $\theta$, and roughness, $\Delta$, in the framework of the bilayer model\cite{ref10,ref17,ref18}.  We begin with the general expression for the RSI ${I(t)}$\cite{ref10,ref17}, ${I(t)=I_{0}[1-2\theta_{1}+2\theta_{2}]^{2}}$, where $\theta_{1}$ and $\theta_{2}$ are the time-dependent coverages of the 1st and the 2nd atomic layers, respectively, and ${I_{0}}$ is a scaling factor. In addition, the coverage is constrained to ${\theta_{1}+\theta_{2}=mt}$, where $m$ is the deposition rate, which vanishes as the laser is switched off (i.e. ${m =0}$ at $ {t>t_{E}}$). Next, we set  ${mt_{E} =1}$ implying that $\theta_{2}$, if present during the deposition, should vanish during the dwell time. The corresponding roughness can be calculated as ${\Delta(t)={\sqrt{\theta_{1}(1-\theta_{1}-\theta_{2})^{2}+\theta_{2}(2-\theta_{1}-\theta_{2})^{2}}}}$. Under these conditions, the experimental RSI have been used to obtain $\theta_{1}$, $\theta_{2}$ and $\Delta$ as a function of time and temperature. The results of the numerical treatment are presented in Figs. 3 (a, b). As anticipated for all the temperatures, the coverage of the 1st layer scales with the deposition time and is clearly T-independent during stage 1 (see Fig. 3(a)). On the other hand, $\theta_{2}$ shows T-dependence only as the RSI approaches its lowest value at ${t_{M}}$. In addition, the roughness, $\Delta$ shown in Fig. 3(b) increases rapidly from the start of deposition. Based on these observations, we infer that during stage 1 the \textit{kinetics-driven condensation} is dominant over the substrate temperature and is crucial for the high-quality LBL growth.

Starting from the point $t_M$, the data obtained between 450$^{o}$C and 780$^{o}$C are markedly T-dependent. The SLs grown at around 700$^{o}$C and above continue the same tendency found in stage 1, namely the continuing increase of  $\theta_{1}$ accompanied by the rapid decline of $\theta_{2}$. In sharp contrast, for the samples grown at 450$^{o}$C and 580$^{o}$C, $\theta_{1}$ becomes progressively slower, and instead $\theta_{2}$ continues its strong progression. This behavior attests that the temperature-controlled diffusion becomes a key factor to maintain 2D growth and to suppress the development of 3D islands inherent to the colder deposition. The strong T-dependence of $\theta_{2}$ also implies that for high temperatures the occasionally grown 2nd level is able to rapidly relax to the 1st level, whereas for colder deposition this process is strongly suppressed resulting in the rapidly increasing roughness. Additionally, strong evidence for the new scaling law is deduced from an analysis of the T-dependence of the RSI after $t_{M}$. Indeed, the T-independent scaling of stage 1 has now switched to a power law with exponent exponent $\beta\approx$1/4 (further details in auxiliary supplement\cite{ref11}). This marked difference can be further explicated from the observation that at $t_{M}$, $\theta_{1}\approx$0.5. This specific value is known to be critical in the 2D percolation problem\cite{ref12}. At this point the new spatial arrangement appears and is characterized by a developing long-range order parameter as experimentally evidenced by the recovering RSI and characteristic morphology of percolated islands shown in Fig 2(b).

Finally, upon completion of the ablation at $t=t_{E}$ (stage 3 in Fig. 3(a)), the temporal dependence of the coverage changes again. During this relaxation period, $\theta_{2}$ starts decreasing,  since the 1st layer recovers at the expense of the 2nd layer. However, this recovery is strongly T-dependent. As seen, for the low temperature deposition $\theta_{1}$ cannot recover in full, whereas at 700$^{o}$C the recovery is practically complete. During this phase the disorder configuration undergoes yet another change towards a new power law with $\beta\approx$1. The experimental observation of the scaling specific to a particular mechanism of growth is important, because during each stage the system evolves by changing only a mean size of the surface disorder and without switching the disorder distribution function defined by the growth mode\cite{ref19,ref20}.

In summary, we have developed the LBL growth of  a novel 1u LNO/1uc LAO HS on STO substrate with excellent morphological quality. We have demonstrated the profound difference between the kinetics of continuous and interrupted growth of this HS and have investigated different stages of sub-monolayer nucleation and growth at the limit of high supersaturation and rapid flux modulation. These experimental findings are of fundamental and practical importance and should enable synthesis of unusual metastable material phases of novel ultra-thin complex oxide heterostructures.

Acknowledgements.

J.C. was supported by DOD-ARO under the Grant No. 0402-17291 and NSF Grant No. DMR-0747808.

\newpage

\begin{figure}[t]\vspace{-0pt}
\includegraphics[scale=.5]{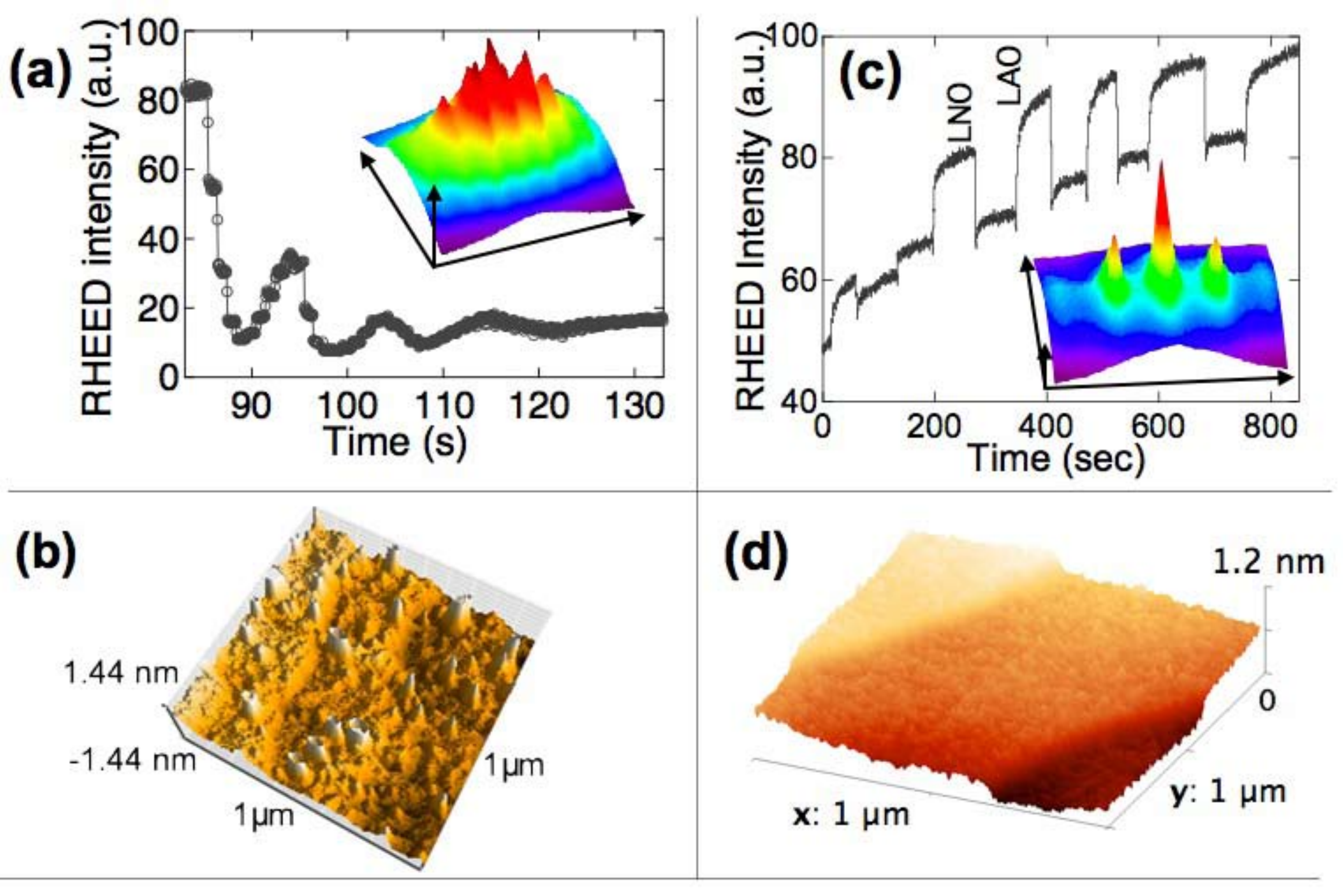}
\caption{\label{Figure1}(Color online) (a) RSI oscillations of LNO/LAO SL (n$\approx$6) grown at ${T=730^{o}}$C, ${P_{O_{2}}}\sim$100 mTorr and with a laser power density $\sim$2.2 J/cm$^{2}$.  Continuous deposition with a laser frequency of 3 Hz. (c) Interrupted growth with a frequency of 30 Hz and a dwell time  $\sim$100 sec. Insets: (a,c) RHEED patterns and (b,d) corresponding AFM images.}
\end{figure}

\begin{figure}[t]\vspace{-0pt}
\includegraphics[scale=.7]{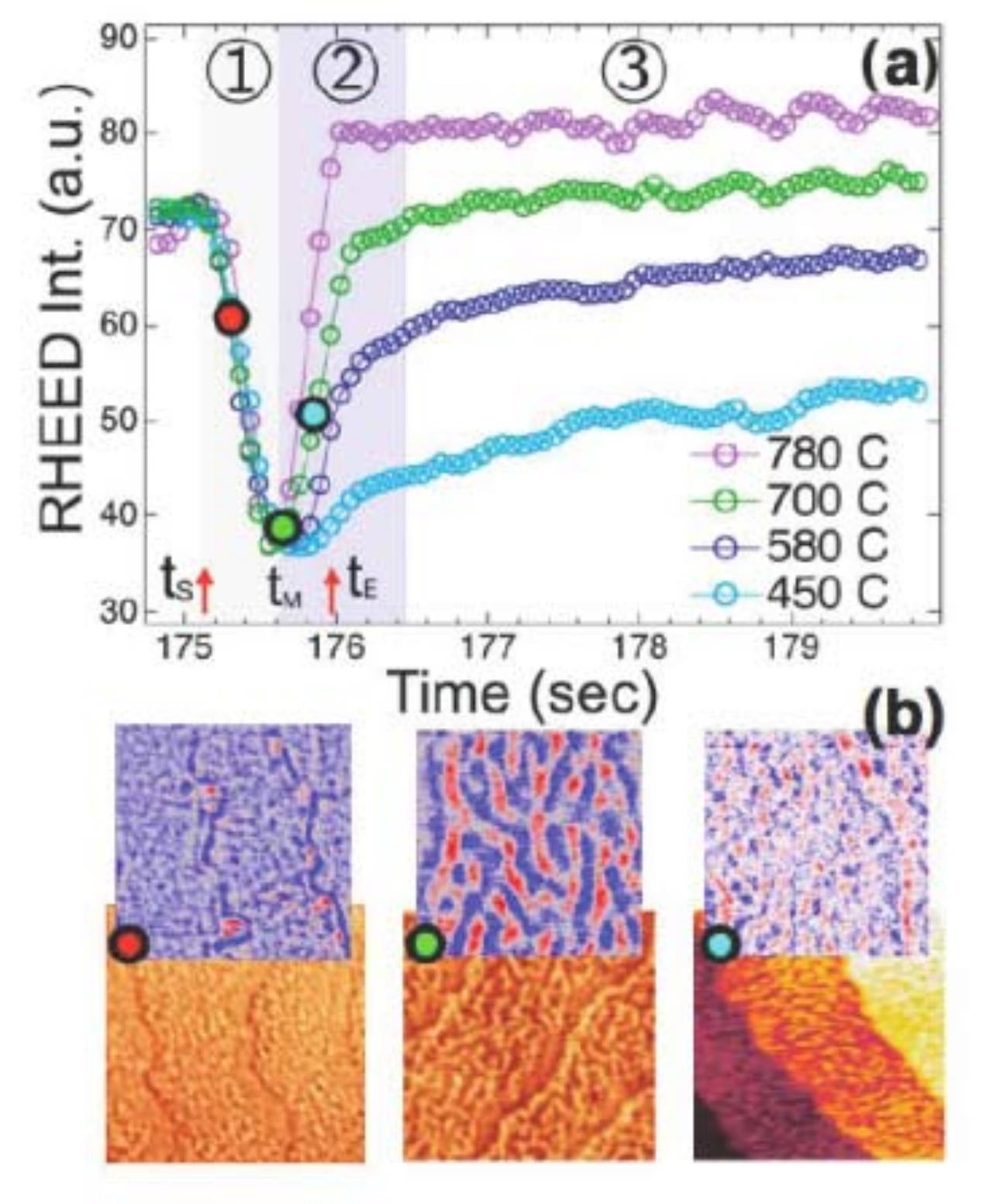}
\caption{\label{Figure2}(Color online) (a) T-dependence of the RSI during the  unit  cell growth. The red arrows indicate the start, $t_{S}$  and the end, $t_{E}$ of the ablation. The lowest diffracted intensity is marked as $t_{M}$. Red, green and blue circles correspond to $\sim$0.3, 0.5 and 0.7 coverage. (b) AFM scans (1 $\times$ 1$\mu$m$^{2}$) obtained by interrupting the growth at the corresponding coverage. Since the evolution of disorder configuration during stage 1 is independent of temperature, the laminae-like  pattern is roughly representative of the same surface seen by RHEED. Insets show the magnified false colored phase-contrast images.}
\end{figure}

\begin{figure}[t]\vspace{-0pt}
\includegraphics[scale=.7]{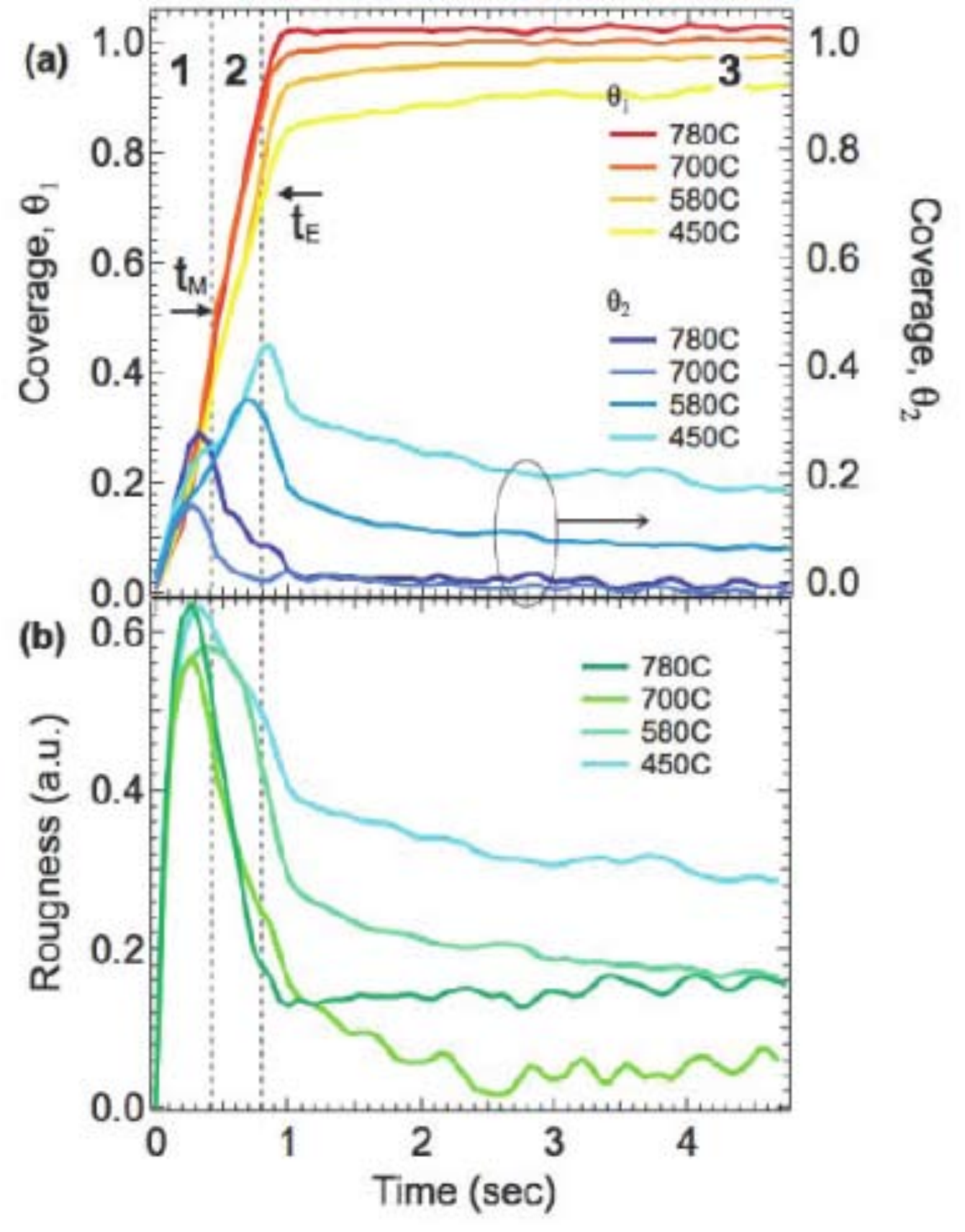}
\caption{\label{Figure3}(Color online) (a) Temperature evolution of surface coverage, $\theta_{1,2}$ and (b) corresponding layer roughness, $\Delta$. The stages of growth marked as 1-3 correspond to the same time-intervals in Fig. 2(a).Note, the difference between the minimum and maximum of RSI is normalized to unity.}
\end{figure}

\begin{figure}[t]\vspace{-0pt}
\includegraphics[width=17cm]{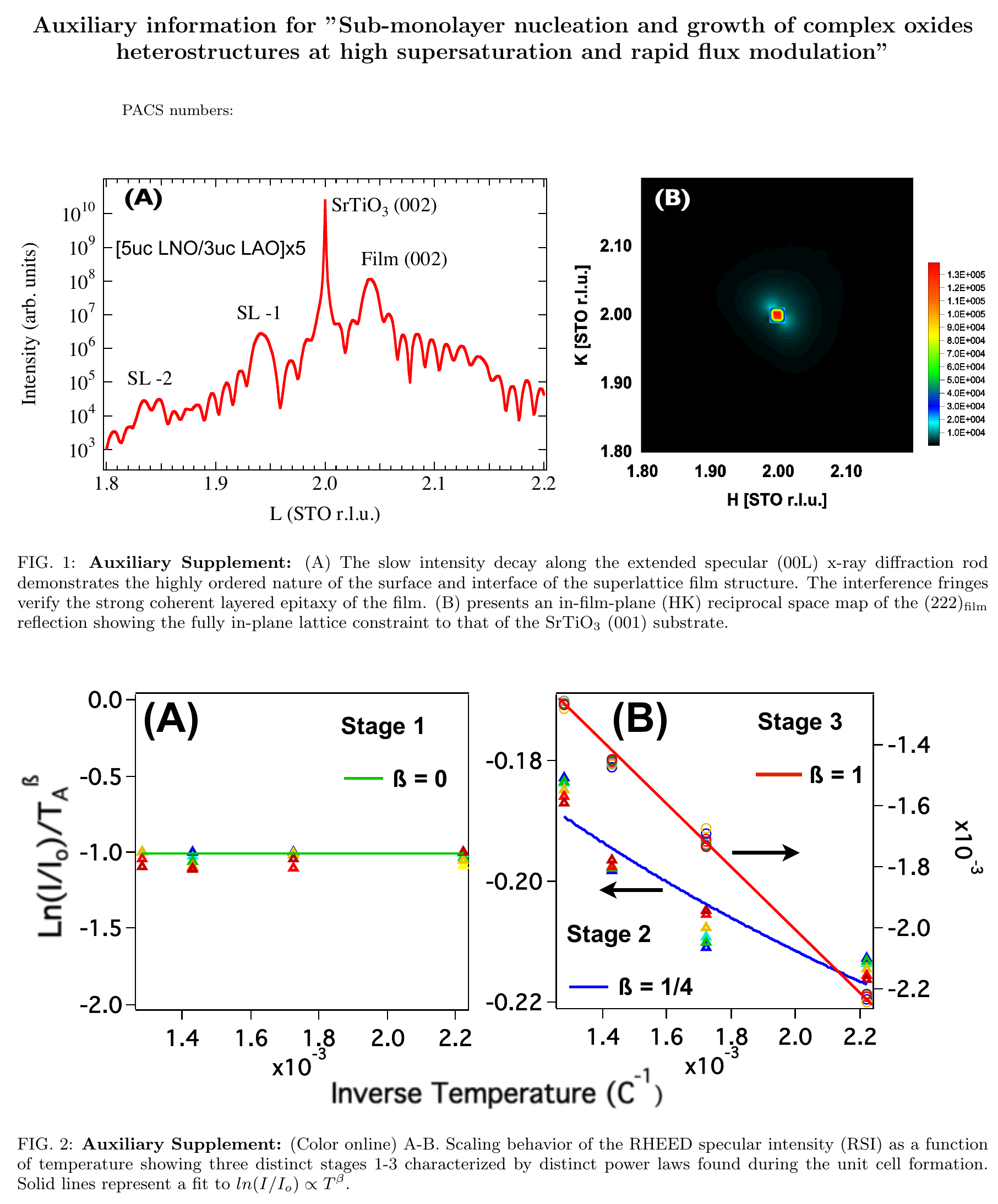}
\end{figure}

\begin{figure}[t]\vspace{-0pt}
\includegraphics[width=17cm]{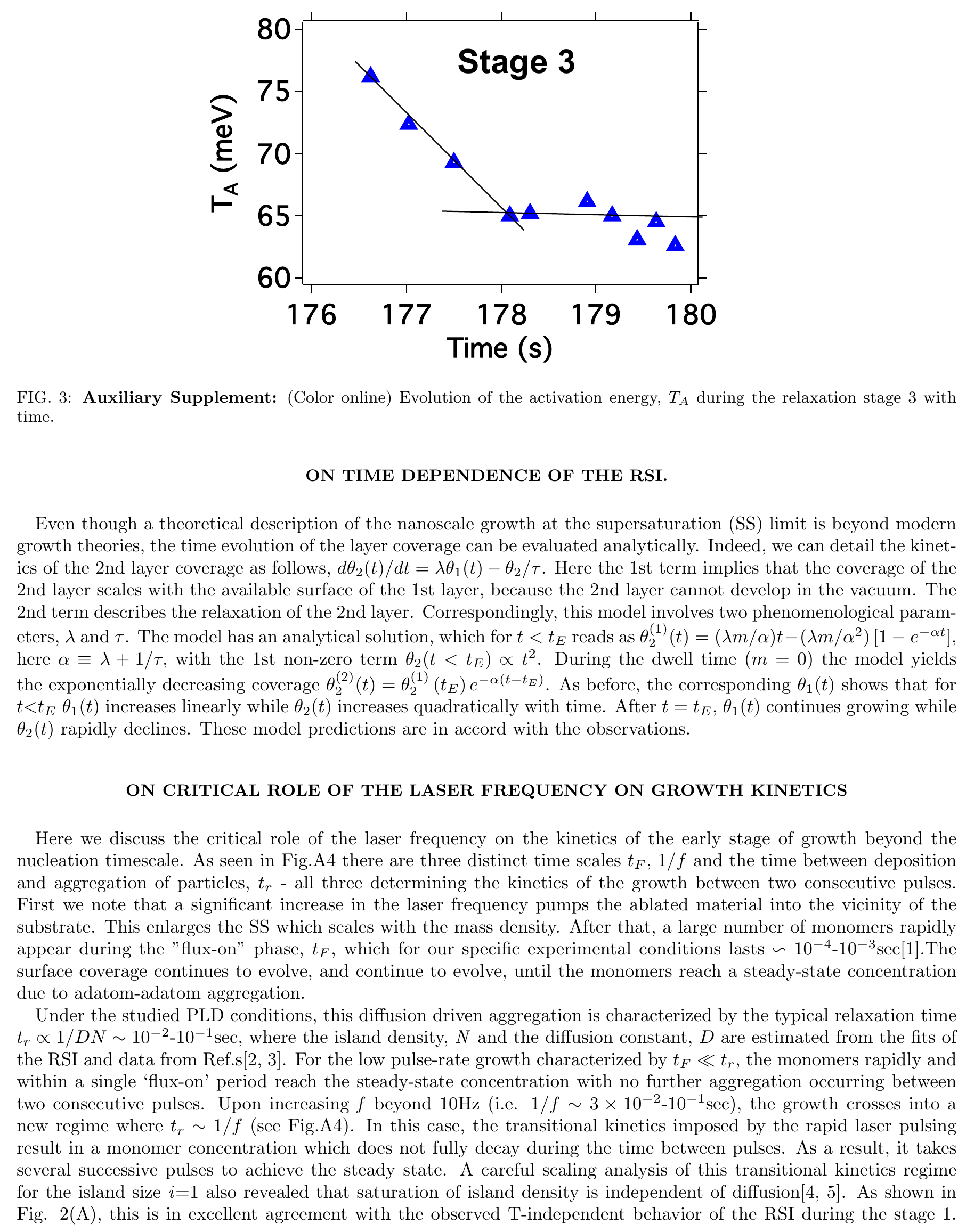}
\end{figure}

\begin{figure}[t]\vspace{-0pt}
\includegraphics[width=17cm]{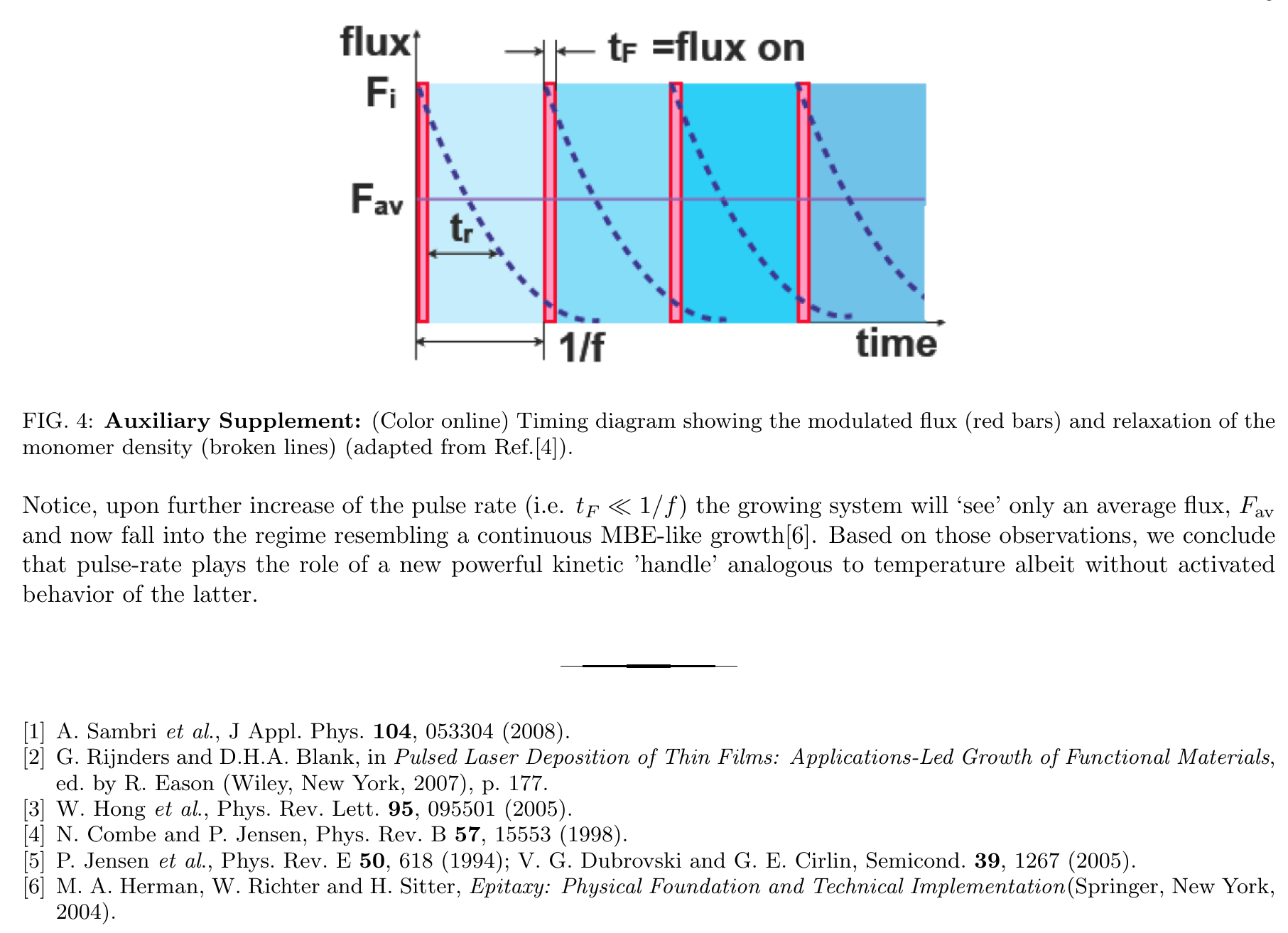}
\end{figure}

\end{document}